# Whole Heart Anatomical Refinement from CCTA using Extrapolation and Parcellation

Hao Xu[1], Steven A. Niederer[1], Steven E. Williams[1,2], David E. Newby[2], Michelle C. Williams[2], and Alistair A. Young[1]

[1] Department of Bioengineering, King's College London, London, UK
[2] University/BHF Centre for Cardiovascular Science, University of Edinburgh, Scotland
`hao2.xu@kcl.ac.uk`

**Abstract.** Coronary computed tomography angiography (CCTA) provides detailed anatomical information on all chambers of the heart. Existing segmentation tools can label the gross anatomy, but addition of application-specific labels can require detailed and often manual refinement. We developed a U-Net based framework to i) extrapolate a new label from existing labels, and ii) parcellate one label into multiple labels, both using label-to-label mapping, to create a desired segmentation that could then be learnt directly from the image (image- to-label mapping). This approach only required manual correction in a small subset of cases (80 for extrapolation, 50 for parcellation, compared with 260 for initial labels). An initial 6-label segmentation (left ventricle, left ventricular myocardium, right ventricle, left atrium, right atrium and aorta) was refined to a 10-label segmentation that added a label for the pulmonary artery and divided the left atrium label into body, left and right veins and appendage components. The final method was tested using 30 cases, 10 each from Philips, Siemens and Toshiba scanners. In addition to the new labels, the median Dice scores were improved for all the initial 6 labels to be above 95% in the 10-label segmentation, e.g. from 91% to 97% for the left atrium body and from 92% to 96% for the right ventricle. This method provides a simple framework for flexible refinement of anatomical labels. The code and executables are available at cemrg.com.

## 1 Introduction

Coronary computed tomography angiography (CCTA) is a widely used imaging tool for investigation of the coronary artery anatomy in patients with suspected coronary artery disease [1]. However, a lot of anatomical information is also present in these scans [2]. Although CCTA has high signal and spatial resolution relative to MRI or echocardiography, existing segmentation methods are difficult to adapt to different applications. In particular, planning ablation therapy for atrial fibrillation requires the identification of left and right pulmonary veins (LPV, RPV) and their intersection with the left atrium (LA) body. Similarly, anatomical features such as the pulmonary artery valve (PAV) and left atrial appendage (LAA) are important for particular pathologies (such as tetralogy of Fallot and stroke respectively).

**Previous work.** Neural network whole heart segmentation methods have previously shown good results with CCTA data [3]. In particular, Baskaran *et al.* [4] applied a 2D U-Net, using 132 training, 34 validation cases and 17 test cases, to predict 5 labels: left ventricle (LV), right ventricle (RV), LA, right atrium (RA) and LV myocardium (LVMyo). Median Dice scores ranged from 0.915 (RV) to 0.938 (LV). LPV and RPV were excluded from the LA but the LAA was included. In the 2017 Multi-Modality Whole Heart Segmentation (MMWHS) challenge, a variety of methods performed well on CCTA datasets (n=60) with 7 labels: LV, RV, LA (excluding LPV, RPV and LAA), LVMyo, ascending aorta (AA), and pulmonary artery (PA) [3]. The leading method used a two-step process, with a localization 3D U-Net and heatmap regression and a subsequent 3D U-Net for segmentation [5].

In this paper, we present a method to adapt algorithms to a different label definition, leveraging existing segmentation tools derived from different sources. We apply our method to the problem of PAV localization and LA parcellation into LPV, RPV, LAA and LA body segments.

## 2      Methods

We describe a multi-stage process (Figure 1), in which existing segmentations were used to provide ground truth for an initial image-to-label 3D U-Net (*U-Net 1*) using 200/30/30 (train/validation/test) cases, giving 6 regions directly from CCTA images. We then refined the segmentation and manually identified the PAV to separate PA from RV, and the image-to-label network was retrained with the refined 6 labels (*U-Net 1 no PA*, 200/30/30 cases). A label-to-label 3D U-Net (*U-Net 2*) was trained to extrapolate the PA label (62/9/9 cases), and another label-to label 3D U-Net (*U-Net 3*) was trained to parcellate LA into LA body, LPV, RPV and LAA (38/6/6 cases). The resulting networks were applied to 1770 cases, and results from 260 cases were manually reviewed and used to train and test a final image-to-label 3D U-Net *(U-Net 4*, 200/30/30) to predict the refined labels directly from the CCTA images.

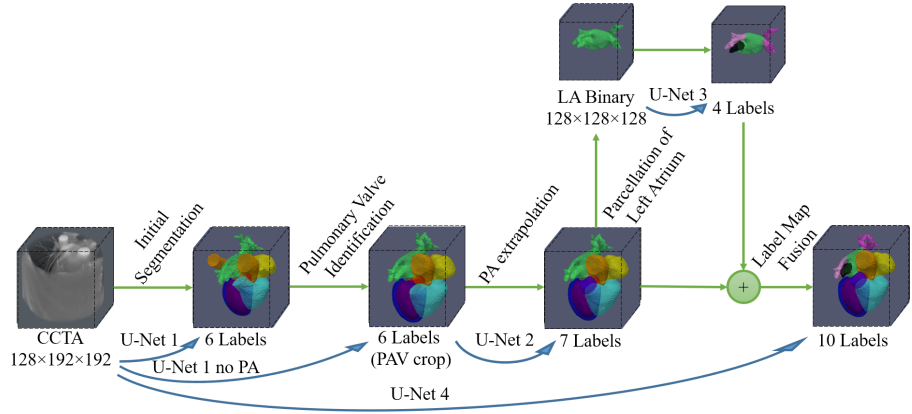

**Fig. 1.** Refinement approach overview. *U-Net 1*, *U-Net 1 no PA*, and *U-Net 4* are image to label maps; *U-Net 2* is a label-to-label extrapolation map; *U-Net 3* is a label-to-label parcellation map.

### 2.1 Data

CCTA exams of 1770 patients (56% male, 58±10 years old) who participated in the Scottish COmputed Tomography of the Heart (SCOT-HEART) trial were included in this study; patient demographics have been reported previously [6]. Briefly, all patients had suspected angina attributable to coronary artery disease and were imaged between 2010 and 2014 at one of three sites using either 64- or 320-detector row scanners (Brilliance 64, Philips Medical Systems, Netherlands; Biograph mCT, Siemens, Germany; Aquilion ONE, Toshiba Medical Systems, Japan). Tube current, voltage, and volume of iodine-based contrast were adjusted based on body mass index. To illustrate how different sources of data can be combined, we also included 20 cases from the MMWHS challenge training dataset and 40 cases from the MMWHS challenge testing dataset [3]. These manually annotated cases were obtained from two 64-slice scanners (both Philips) at two sites in Shanghai, China.

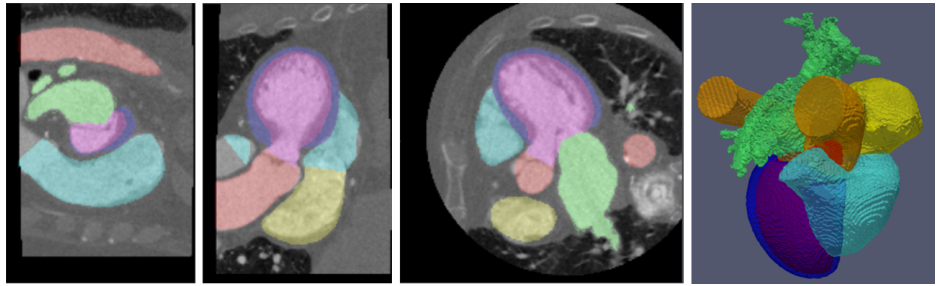

**Fig. 2.** Coronal, sagittal and axial views of the CCTA image overlapped with initial segmentation labels, and 3D visualization of the anatomies. LV, LVMyo, RV, LA, RA and AA are in purple, dark blue, light blue, green, yellow and orange respectively.

### 2.2 Image-to-Label Initial Segmentation

For the SCOT-HEART cases an initial segmentation was automatically performed using Siemens AXseg v4.11 prototypical software (Siemens Healthineers, Erlangen, Germany). This method used an atlas combined with marginal space learning and steerable filters [7]. Six regions were labelled: LV, LVMyo, RV, LA, RA, and AA. For the MMWHS cases the manual annotations provided for the corresponding regions were used. The image and label maps were normalized to voxel size of 1 mm$^3$ and cropped or padded to volumes with size of 128x192x192 with the heart at the center of the volume. For large hearts, the voxel size was iteratively increased by 10% until the field of view covered all labelled voxels. Typical results of the initial segmentation are shown in Figure 2.

A 3D U-Net (*U-Net 1*) was designed with 3 max-pooling and deconvolutional stages with a stride of 2x2x2. The numbers of convolutional kernels were set to be (16, 32), (32, 64), (64,128) for the contraction path and (128, 256) for the bottle neck, with the kernel size of 3x3x3. The numbers of deconvolutional kernels were set to be (128, 128), (64, 64), (32, N) for the expansion path, where N is the number of layers of the output volume. For the initial segmentation network N=7 (background and 6 labels). The network was trained with cross-entropy loss, using 200 training (180 randomly selected from SCOT-HEART and 20 from MMWHS) cases, with additional 30 cases for validation and 30 cases for testing. The training, validating and testing cases from SCOT-HEART were randomly and equally sampled from three types of scanner. Dice scores were used for evaluation.

### 2.3 Label Processing

We found that several refinements were necessary to improve the accuracy and consistency of the initial segmentations. Firstly, AXseg tended to visually over-segment the LV cavity in non-Siemens scanners (Figure 2). We therefore used a morphological operator to dilate the LVMyo mask, and voxels overlapping between dilated LVMyo and LV were transferred to the LVMyo label if i) the mean intensity value of the overlap region was closer to LVmyo than LV, and ii) the voxel intensity was less than the mean plus one standard deviation of the original LVMyo voxel intensities. This process was repeated up to a maximum of three times.

Secondly, a PA label was used for PAV identification, with the PAV defined as the intersection of PA with the RV (all voxels in the RV with a PA neighbor, and vice versa). This method was preferred to the direct segmentation of the PAV, since U-Nets do not work as well on classes with such small number of voxels due to class-imbalance. However, a significant number of SCOT-HEART cases cropped the PA from the field of view, resulting in inconsistent PA segmentations. Furthermore, the AXseg tool did not include a PA label but included a PA section (if present) in the RV label. Therefore, we manually partitioned the initial RV label for the 260 cases used to train and test *U-Net 1* into RV and PA labels using a PAV plane defined using landmarks on the reformatted images. The initial image-to-label network was then retrained without the PA section, using the refined RV label (*U-Net 1 no PA* in Figure 1).

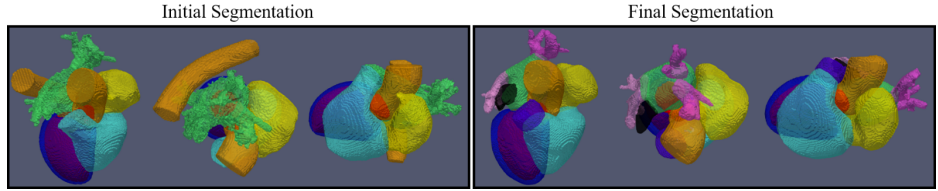

**Fig. 3.** Examples of image to label results for initial (6 labels) and final (10 labels) maps.

### 2.4 Label to Label Networks

*Extrapolation of PA*
In order to enable prediction of a PA label by extrapolating the RV outflow tract, even in cases in which the PA was cropped, we chose 80 cases with PA present (from the original 260 cases used to train *U-Net 1 no PA*, with ground truth partitioned into RV and PA) split into 62 training cases, 9 validation cases and 9 testing cases. We applied same 3D U-Net architecture (section 2.2) to map predictions from *U-Net 1 no PA* to a 7-label segmentation with PA included (*U-Net 2*). The input/output volumes were set to be 128x192x192.

*LA Parcellation*
We manually partitioned the LA into body, LPV, RPV and LAA labels (multiple LPV were given the same label, similarly for RPV) using 3DSlicer's cropping boxes (scissors tool). Compared to PA extrapolation, manual annotation was more time-consuming, but the number of cases requiring annotation was smaller as it is simpler for a label-to-label network to learn how to relabel an existing structure (initial LA label) than to predict a new one. Therefore, only 50 cases were required (38 training cases, 6 validating cases and 6 testing cases). The same 3D U-Net architecture (section 2.2) was applied and input/output volumes were set to be 128x128x128 (*U-Net 3*).

### 2.5 Image to Label Refinement

The image-to-label 3D U-Net was retrained to directly predict 10-label maps from CCTA input (*U-Net 4*). The ground truth for this network was generated by applying the refined initial segmentation network and label-to-label networks, fusing the predictions, and manually evaluating the result using ITK-snap, until 260 predictions with good quality were identified. The segmentations consisting of one object were cleaned by choosing the largest connected component. This process resulted in 200 training (180 from SCOT-HEART and 20 from MMWHS) cases, with additional 30 validation cases and 30 test cases, evenly split by scanner types. Dice scores were used for evaluation against the output of the label-to-label network results. An example of a pair of initial and refined segmentations are shown in Figure 3.

## 3 Results

### 3.1 Image to Label Initial Segmentation

Dice scores for *U-Net 1* are shown in Figure 4 for the testing dataset of 30 cases. The network performed well on RV, LA and RA cavities, with median Dice scores above 90%. The network performed less well on LV cavity, LVMyo and AA, with lower median values and higher variations of the Dice scores. For LV and LVMyo, the Dice scores reflect a larger variation around the LV endocardial surface across scanner types, resulting in both a reduction in the overall similarity and a larger variation of the Dice scores. The LVMyo volume was smaller and therefore showed lower Dice than LV cavity. The larger variations of the RV and AA scores reflect the effect of PA and descending aorta variations.

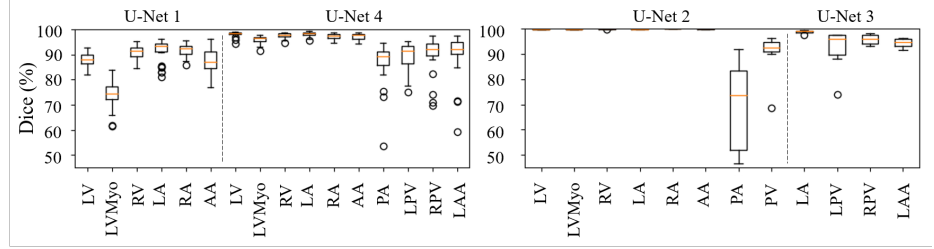

**Fig. 4.** Dice scores for test cases of image-to-label networks (*U-Net 1* and *U-Net 4*, n=30) and label-to-label networks (*U-Net 2* n=9, and *U-Net 3*, n=6).

### 3.2 Label to Label Refinement

Dice scores for the label-to-label U-Nets (*U-Net 2* and *U-Net 3*) are shown in Figure 4. *U-Net 2* extrapolated from 6 labels to 7 labels, and the 6 input labels were almost identically reproduced by the network with Dice scores all >99%. The PA Dice scores were relatively lower reflecting the large variation of PA cropping. However, the goal of the PA label was to identify the PAV, and the resulting PAV Dice was very good considering it has a thickness of just two voxels. The outliers of PAV Dice in *U-Net 2* represent cases with incomplete PAV caused by the limited scanning field of view, which was common within SCOT-HEART. The initial segmentations were constrained by the image size, however, the extrapolation approach mitigated this problem by extrapolating the RV outflow tract as shown in Figure 5.

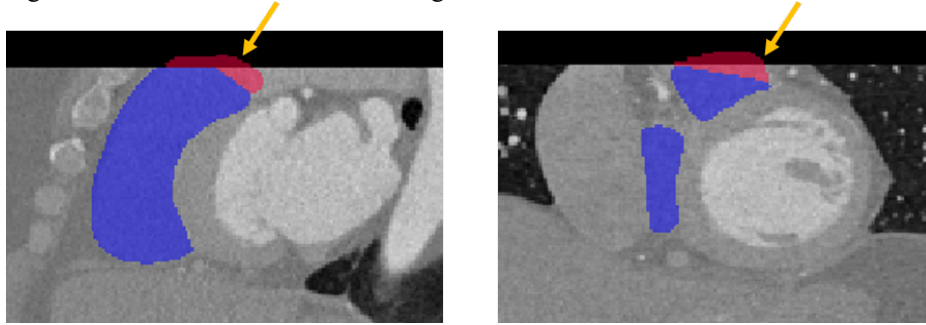

**Fig. 5.** Coronal and sagittal views of the CCTA image overlapped with reconstructed RV (blue) and PA (red) labels from U-Net 2 predictions for the outlier testing case. The yellow arrows indicate the predicted PA label voxels outside the scanning field of view.

*U-Net 3* mapped the initial LA label to LA body, LPV, RPV and LAA labels. The Dice scores showed good similarity between the reference and the predicted labels, especially for LA body. The Dice score was affected by the size of the object, and therefore the Dice scores for LPV, RPV and LAA were relatively lower.

### 3.3 Image to Label Final Segmentation

Dice scores for *U-Net 4*, calculated against the 10-label predictions for the manually reviewed test cases, are shown in Figure 4. The median Dice scores are all above 95% for 6 initial labels, with the largest value above 98% for LV cavity. The mean values of Dice scores are also between 95% and 98% with the standard deviations between 1% and 2%. The reduction of outliers suggests that the consistency of the segmentations across scanner types was clearly improved through our refinement pipeline, including LV cavity adjustment, PA extrapolation and LA parcellation. The final segmentation network could also accurately predict the refined labels from CCTA images showing the manual annotations were highly correlated to the image features. As expected, the Dice scores for small regions are not as good as big regions, and PA has the lowest median Dice score (89%) while LPV, RPV and LAA all have median Dice score above 90%. The larger variations and more outliers reflected images with partially cropped field of view superior to the LV. Compared with previous reports [3, 4, 8, 9], our results show high performance even in smaller regions. We also evaluated *U-Net 1* and *U-Net 4* using the testing dataset of the MMWHS challenge, and compared to the top-5 ranked participants of the challenge for CT image segmentation [10]. The Dice scores are shown in Table 1. The Dice score of *U-Net 1* is smaller because it is trained using automatically generated segmentation as the reference, and there is an obvious improvement after our refinement process, giving the performance of *U-Net 4* similar to the best performing challenge participants.

Table 1. Dice on the CT test datasets of the MMWHS challenge for U-Net 1, U-Net 4, and the top-5 ranked participants. The values are the mean of 40 cases in %.

|         | LV   | LVMyo | RV   | LA   | RA   |
|---------|------|-------|------|------|------|
| U-Net 1 | 88.0 | 81.5  | 83.4 | 81.6 | 82.2 |
| U-Net 4 | 90.1 | 84.7  | 89.2 | 91.7 | 87.7 |
| 1       | 91.8 | 88.1  | 90.9 | 92.9 | 88.8 |
| 2       | 92.3 | 85.6  | 85.7 | 93.0 | 87.1 |
| 3       | 90.4 | 85.1  | 88.3 | 91.6 | 83.6 |
| 4       | 90.1 | 84.6  | 85.6 | 88.4 | 83.7 |
| 5       | 90.8 | 87.4  | 80.6 | 90.8 | 85.5 |

## 4 Conclusions and Limitations

In this paper we described a flexible method for refining cardiac segmentations using a multi-stage process, by training both image-to-label and label-to label networks to learn task-dependent manual corrections in an iterative fashion. The label-to-label networks were able to learn the anatomical configuration of the refined label maps using only mask information by either extrapolating new labels from background (U-Net 2) or partitioning existing labels (U-Net 3). Taking the advantage of neural networks, which provided a simple way to refine or adapt segmentations to suit the application, we applied this method to the problem of identifying the PAV, as well as the locations of the pulmonary veins and LAA with the LA.

Limitations of this study include i) no cases had metal artefacts which are often present in CCTA, and ii) the PAV could be distorted in some cases due to restricted field of view (Figure 5) – this could be corrected in future work by manually extrapolating the RV in addition to the LA for subsequent label-to-label refinement.

**Acknowledgements.** This research was supported by the UKRI London Medical Imaging & Artificial Intelligence Centre for Value Based Healthcare, and core funding from the Wellcome/EPSRC Centre for Medical Engineering [WT203148/Z/16/Z]. We thank Siemens Healthineers for allowing us to use the AXseg prototypical software during this research. SCOT-HEART was funded by The Chief Scientist Office of the Scottish Government Health and Social Care Directorates (CZH/4/588), with supplementary awards from Edinburgh and Lothian's Health Foundation Trust and the Heart Diseases Research Fund. MCW and DEN are supported by the British Heart Foundation FS/ICRF/20/26002 and CH/09/002.

**References**

1. Knuuti J, Wijns W, Saraste A, Capodanno D, Barbato E, Funck-Brentano C, Prescott E, Storey RF, Deaton C, Cuisset T *et al*: 2019 ESC Guidelines for the diagnosis and management of chronic coronary syndromes. *Eur Heart J* 2020, 41(3):407-477.
2. Hoogendoorn C, Duchateau N, Sanchez-Quintana D, Whitmarsh T, Sukno FM, De Craene M, Lekadir K, Frangi AF: A high-resolution atlas and statistical model of the human heart from multislice CT. *IEEE Trans Med Imaging* 2013, 32(1):28-44.
3. Zhuang X, Li L, Payer C, Stern D, Urschler M, Heinrich MP, Oster J, Wang C, Smedby O, Bian C *et al*: Evaluation of algorithms for Multi-Modality Whole Heart Segmentation: An open-access grand challenge. *Med Image Anal* 2019, 58:101537.
4. Baskaran L, Maliakal G, Al'Aref SJ, Singh G, Xu Z, Michalak K, Dolan K, Gianni U, van Rosendael A, van den Hoogen I *et al*: Identification and Quantification of Cardiovascular Structures From CCTA: An End-to-End, Rapid, Pixel-Wise, Deep-Learning Method. *JACC Cardiovasc Imaging* 2020, 13(5):1163-1171.
5. Payer C, Stern D, Bischof H, Urschler M: Multi-label Whole Heart Segmentation Using CNNs and Anatomical Label Configurations. In: *Statistical Atlases and Computational Models of the Heart ACDC and MMWHS Challenges STACOM 2017 LNCS. Volume 10663*, edn.: Springer; 2018.
6. Williams MC, Kwiecinski J, Doris M, McElhinney P, D'Souza MS, Cadet S, Adamson PD, Moss AJ, Alam S, Hunter A *et al*: Low-Attenuation Noncalcified Plaque on Coronary Computed Tomography Angiography Predicts Myocardial Infarction: Results From the Multicenter SCOT-HEART Trial (Scottish Computed Tomography of the HEART). *Circulation* 2020, 141(18):1452-1462.
7. Zheng Y, Barbu A, Georgescu B, Scheuering M, Comaniciu D: Four-chamber heart modeling and automatic segmentation for 3-D cardiac CT volumes using marginal space learning and steerable features. *IEEE Trans Med Imaging* 2008, 27(11):1668-1681.

8. Leventić H, Babin D, Velicki L, Devos D, Galić I, Zlokolica V, Romić K, Pižurica A. Left atrial appendage segmentation from 3D CCTA images for occluder placement procedure. *Computers in biology and medicine*. 2019 Jan 1;104:163-74.
9. Jin C, Feng J, Wang L, Yu H, Liu J, Lu J, Zhou J. Left atrial appendage segmentation using fully convolutional neural networks and modified three-dimensional conditional random fields. *IEEE journal of biomedical and health informatics*. 2018 Jan 17;22(6):1906-16.
10. Payer, C., Štern, D., Bischof, H. and Urschler, M., 2017, September. Multi-label whole heart segmentation using CNNs and anatomical label configurations. In International Workshop on Statistical Atlases and Computational Models of the Heart (pp. 190-198). Springer, Cham.